\documentclass[emulateapj,natbib]{emulateapj}
\begin{document}

%\title{Gemini Planet Imager Imaging and Polarimetry of HD 100546: H band Recovery of HD 100546 b and Detection of A Candidate Second Protoplanet or Bright, Weakly Polarized Inner Disk Wall}
%\title{Gemini Planet Imager Near-Infrared Photometry of HD 100546 b with the Gemini Planet Imager, Recovery of Thermal-IR Bright Spiral Arm, and Detection of A Candidate Second Protoplanet (or Bright Inner Disk Wall)}
%\title{Gemini Planet Imager Detection of the Young Protoplanet HD 100546 b and Candidate Second Protoplanet or Weakly-Polarized, Bright Disk Wall at 13 AU}
%\title{Resolving the  HD 100546 Protoplanetary System with the \textrm{Gemini Planet Imager}: Evidence for Multiple Forming, Accreting Planets}
%\textit{A First \textrm{Gemini Planet Imager} Study of the HD 100546 Protoplanetary System: Recovery of HD 100546 b
%\title{Direct Imaging of a Thermal IR Bright, Solar System-Scale Inner Disk Rim In Scattered Light Around HD 141569A}
\title{The Matryoshka Disk: Keck/NIRC2 Discovery of a Solar System-Scale, Radially Segregated Residual Protoplanetary Disk Around HD 141569A}
%\footnote{As in a ``matryoshka doll" or a Russian nesting doll}
\author{
Thayne Currie\altaffilmark{1}, 
Carol A. Grady\altaffilmark{2},
Ryan Cloutier\altaffilmark{3},
Mihoko Konishi\altaffilmark{4},
Keivan Stassun\altaffilmark{5},
John Debes\altaffilmark{6},
Nienke van der Marel\altaffilmark{7},
Takayuki Muto\altaffilmark{8},
%Sebastian Daemgen\altaffilmark{8},
Ray Jayawardhana\altaffilmark{9},
Thorsten Ratzka\altaffilmark{10}
%Adam Burrows\altaffilmark{4},
%Marc Kuchner\altaffilmark{5}
%,
%Ryan Cloutier\altaffilmark{8}
}
\altaffiltext{1}{National Astronomical Observatory of Japan, Subaru Telescope, Hilo, Hawaii}
%\altaffiltext{3}{Kogakuin University}
%\altaffiltext{4}{Department of Astrophysics Sciences, Princeton University}
\altaffiltext{2}{Exoplanets and Stellar Astrophysics Laboratory, NASA-Goddard Space Flight Center, Greenbelt, Maryland}
\altaffiltext{3}{Department of Astronomy and Astrophysics, University of Toronto, Toronto, Canada}
\altaffiltext{4}{Department of Earth and Space Sciences, Graduate School of Science, Osaka University, Osaka, Japan}
\altaffiltext{5}{Department of Physics and Astronomy, Vanderbilt University, Nashville, Tennessee}
\altaffiltext{6}{Space Telescope Science Institute, Baltimore, Maryland}
\altaffiltext{7}{Institute for Astronomy, University of Hawaii-Manoa}
\altaffiltext{8}{Kogakuin University}
%\altaffiltext{8}{ETH-Zurich, Zurich, Switzerland}
\altaffiltext{9}{Department of Physics and Astronomy, York University, Toronto, Canada}
\altaffiltext{10}{Institute for Physics/IGAM, NAWI Graz, University of Graz, Graz, Austria}
%\altaffiltext{10}{Universitats-Sternwarte Munchen, Ludwig-Maximilians-Universitat, Scheinerstr. 1, 81679, Munchen, Germany}
%\altaffiltext{8}{Department of Astronomy and Astrophysics, University of Toronto}
%\altaffiltext{4}{European Southern Observatory}
%\altaffiltext{2}{University of Tokyo}
%\altaffiltext{3}{University of Amsterdam}
\begin{abstract}
Using Keck/NIRC2 $L^\prime$ (3.78 $\mu m$) data, we report the direct imaging discovery of a scattered light-resolved, solar system-scale residual protoplanetary disk around the young A-type star HD 141569A, interior to and concentric with the two ring-like structures at wider separations.   The disk is resolved down to $\sim$ 0\farcs{}25 and appears as an arc-like rim with attached hook-like features.  It is located at an angular separation intermediate between that of warm CO gas identified from spatially-resolved mid-infrared spectroscopy and diffuse dust emission recently discovered with the \textit{Hubble Space Telescope}.  The inner disk has a radius of $\sim$ 39 AU, a position angle consistent with north-up, an inclination of $i$ $\sim$ 56$^{o}$, and has a center offset from the star.  Forward-modeling of the disk favors a thick torus-like emission sharply truncated at separations beyond the torus' photocenter and heavily depleted at smaller separations.   In particular, the best-fit density power law for the dust suggests that the inner disk dust and gas (as probed by CO) are radially segregated, a feature consistent with the dust trapping mechanism inferred from observations of ``canonical" transitional disks.  However, the inner disk component may instead be explained by radiation pressure-induced migration in optically-thin conditions, in contrast to the two stellar companion/planet-influenced ring-like structures at wider separations.   
HD 141569A's circumstellar environment --- with three nested, gapped, concentric dust populations --- is an excellent laboratory for understanding the relationship between planet formation and the evolution of both dust grains and disk architecture.
%We predict that the HD 141569A inner disk could be decisively recovered and well resolved into a torus with upcoming observations, such as mm interferometry using \textit{ALMA}.
%While some reductions identify a candidate point source-like object at the outer edge of the rim, its appearance could instead be due to disk self-subtraction.   We place limits of xx $M{J}$ in between this ring and the previously identified ones.   
%HD 141569A's circumstellar environment exhibits an incredibly resembles a ``hybrid" of different types of disks in different evolutionary phases, where the innermost torus-shaped region resembles that of a residual protoplanetary disk (a transitional disk) whereas the outer regions contain two roughly concentric gas poor debris ring-like structures.
%, and thus an important laboratory for studying the final stages of planet formation.
%At least one newly-forming planet is likely responsible for the gap in the rings: data with the new suite of extreme AO imagers like GPI, SCExAO, and SPHERE could identify it.   
\end{abstract}

\section{Introduction}
Gas-rich, optically thick and luminous protoplanetary disks surrounding young stars typically dissipate in 5--10 $Myr$ by a combination of viscous draining, photoevaporative clearing and/or giant planet formation \citep{Williams2011}.  At older ages, optically-thin gas-poor debris disks whose dust is sustained by planetesimal collisions comprise most of the disk population \citep[][]{KenyonBromley2008,Wyatt2008}.  Directly imaging systems covering the protoplanetary-to-debris disk \textit{transition} reveals a diverse set of disks architectures, probes disk dispersal mechanisms, and identifies evidence for infant jovian planets \citep[e.g.][]{Grady2013,KrausIreland2012,Currie2015b}.

The circumstellar environment around the nearby \citep[$d$ = 116 $pc$][]{vanLeewen2007} A-type star HD 141569A is a particularly good laboratory for studying the last moments of this transition phase.  The system's nominal age \citep[5 $\pm$ 3 $Myr$;][]{Weinberger2000,Aarnio2008} is comparable to the characteristic protoplanetary disk evolution timescale \citep{Cloutier2014}.  While HD 141569 A retains a significant reservoir of gas \citep[$\sim$ 0.2--0.5 $M_{J}$;][]{Zuckerman1995,Thi2014}, its infrared emission is optically thin and its fractional luminosity is little higher than that of luminous debris disk-bearing stars like HR 4796A.  
Near-infrared (IR) to optical scattered-light imaging reveals two nested, bright rings of dust at $r$ $\sim$ 250 and 400 $AU$ exhibiting pericenter offsets and spiral structure driven by  the primary's M dwarf companions and perhaps unseen newly-formed planets \citep{Weinberger1999,Mouillet2001,Clampin2003,Wyatt2005,Janson2013,Biller2015,Mazoyer2016,Konishi2016}.  

 %Data obtained at 10--20 $\mu m$ and recent scattered-light imaging suggest that 
 %HD 141569A's circumstellar environment may be even more complex at solar system like scales ($r$ $<$ 40--50 AU).    
 Additionally, HD 141569A includes warm circumstellar material: CO gas emission cleared out to 11 AU  and marginally resolved warm thermal dust emission (10--20 $\mu m$) potentially depleted interior to 30 AU \citep{Fisher2000,Marsh2002,Goto2006}.  Recently, \citet{Konishi2016} discovered an additional diffuse optical scattered-light component at 40--100 $AU$.  Deeper high-contrast scattered-light imaging may clarify how HD 141569A's circumstellar environment is being cleared of residual protoplanetary material at smaller, solar system-like scales.

In this Letter, we present the discovery of a bright, solar system-scale residual protoplanetary disk around HD 141569A using Keck/NIRC2 $L^\prime$ high-contrast imaging \footnote{We note an independent detection of this inner disk from D. Mawet (Mawet et al. 2016, in prep.)}.   The scattered light-detected dust disk lies interior to the diffuse emission recently discovered by \citet{Konishi2016} but is peaked at radii exterior to the CO gas resolved by \citet{Goto2006}.  The dust disk is likely heavily evacuated at the CO gas's inner radius, revealing evidence for dust/gas segregation.% reminiscent of the planet (dust-) trapping scenario posited for massive, optically thick transitional disks.   
%The disk's presence and torus-like shape could be confirmed by submm interferometry with the \textit{Atacama Large Millimeter Array} (ALMA).

\section{Observations and Data Reduction}
%\subsection{2012 VLT/NaCo Data ($L^\prime$, and [4.05])}
%\subsection{Keck/NIRC2 $L^\prime$ Imaging}

We imaged HD 141569A on 8 June 2015 with the NIRC2 camera on the Keck II telescope on Maunakea in the $L^\prime$ filter ($\lambda_{o}$ = 3.778 $\mu m$) using the narrow camera 
\citep[9.952 mas/pixel;][]{Yelda2010} and the "large hex" pupil plane mask in \textit{angular differential imaging} mode \citep{Marois2006}, and in a 3-point dither pattern (Program N134$\_$N2).  Our science frames consisted of 74 50-second exposures ($t_{int}$ = 0.25 $s$, 200 coadds) for a total integration time of 3700 $s$, 
covering a field rotation of 45.9$^{o}$.   Conditions were photometric 
with slightly above-average quality seeing (0\farcs{}4--0\farcs{}5 in the optical).   We obtained shorter, unsaturated images bracketing our science sequence for flux calibration.

%\subsection{Archival VLT/NaCo $L^\prime$ Imaging}
Additionally, we retrieved archival VLT/NaCo $L^\prime$ imaging of HD 141569A taken on 6 April 2010.   The data were taken in a four-point dither pattern achieving 70.4$^{o}$ of field rotation.  Each of the 292 exposures consists of 150 coadded frames of 0.2 $s$ each for a total integration time of 8760 $s$.   The ESO archival nightly log implies somewhat poorer optical seeing conditions than our Keck data (0\farcs{}65--1\farcs{}0).
%: we use the NaCo data to confirm features seen in the higher-quality Keck/NIRC2 data.  

\subsection{Image Processing}
For basic processing, we utilized the pipeline developed for and followed steps outlined in \citet{Currie2011a,Currie2014a}, sky-subtracting a given frame using the nearest (in time) 
available sky frames and correcting each image for bad pixels.    For the NIRC2 data, we applied a linearity correction.
% \citep[][]{Metchev2009}.
%For the NIRC2 data, we applied the non-linearity correction from \citet{Metchev2009} prior to these steps  
%\footnote{Normally, we also apply the standard NIRC2 distortion solution \citep{Yelda2010}.  However, a new distortion solution is required for data that (like ours) were taken after a May 2015 instrument upgrade.  Applying the previous correction induces astrometric errors: i.e. $\sim$ 10 mas for an object at $r$ $\sim$ 1\arcsec{}.   Omitting the published correction yields smaller astrometric errors.}.  
We improved PSF stability of the NaCo data by realigning each coadd within the cube, removing frames with core-to-halo brightness ratios $\textit{{R}}$ lower than max($\textit{{R}}$) $<$ 3-$\sigma_{\textit{{R}} }$, and collapsing the cube using a robust mean with a 3-$\sigma$ outlier rejection \citep[see][]{Currie2014a}.
As in \citet{Currie2011a}, we registered each image to a common center using a cross-correlation approach.
% and removed frames with a poor PSF quality.
%measured the core to halo brightness ratio  across our image sequence to assess the PSF/AO correction quality and removed poorly corrected frames.  

We performed PSF subtraction using the A-LOCI pipeline \citep{Currie2014c}, an extension and modification of the original locally-optimized combination of images algorithm \citep{Lafreniere2007a}. We use a moving pixel mask \citep{Currie2012} to reduce and normalize throughput, as well as a singular value decomposition (SVD) cutoff  \citep{Marois2010b,Currie2014c}, and speckle filtering/frame selection to reduce errors propagating through the matrix inversion and prevent the solutions from being overdetermined.   
 %Second, we used our implementation of the Kar\'hoeven-Lo\'eve Image Projection (KLIP) algorithm \citep{Soummer2012}.   We construct eigenimages in annular regions, retaining $n_{\rm pca}$ principal components, where $n_{\rm pca}$ $<$ $n_{images}$.  As with A-LOCI, we impose a rotation gap criterion ($\delta$).   As we later found, the disk signal is very strong but at risk for severe self-subtraction and morphological biasing due to its small angular separation.  
 We adopted conservative settings proven successful for detecting bright off-axis signal with only weak algorithm self-subtraction \citep{Currie2015a,Currie2015b}-- i.e.  a large rotation gap equal to the PSF core width ($\delta$ = 1), a high SVD cutoff ($SVD_{lim}$ = 10$^{-3}$), and a large optimization area from which we determine the LOCI coefficients ($N_{A}$ = 1000 PSF footprints).

\section{Detection of the HD 141569A Residual Inner Disk}
\begin{figure}
\centering
%\epsscale{0.8}
\includegraphics[scale=0.4,trim=32mm 1mm 32mm 1mm,clip]{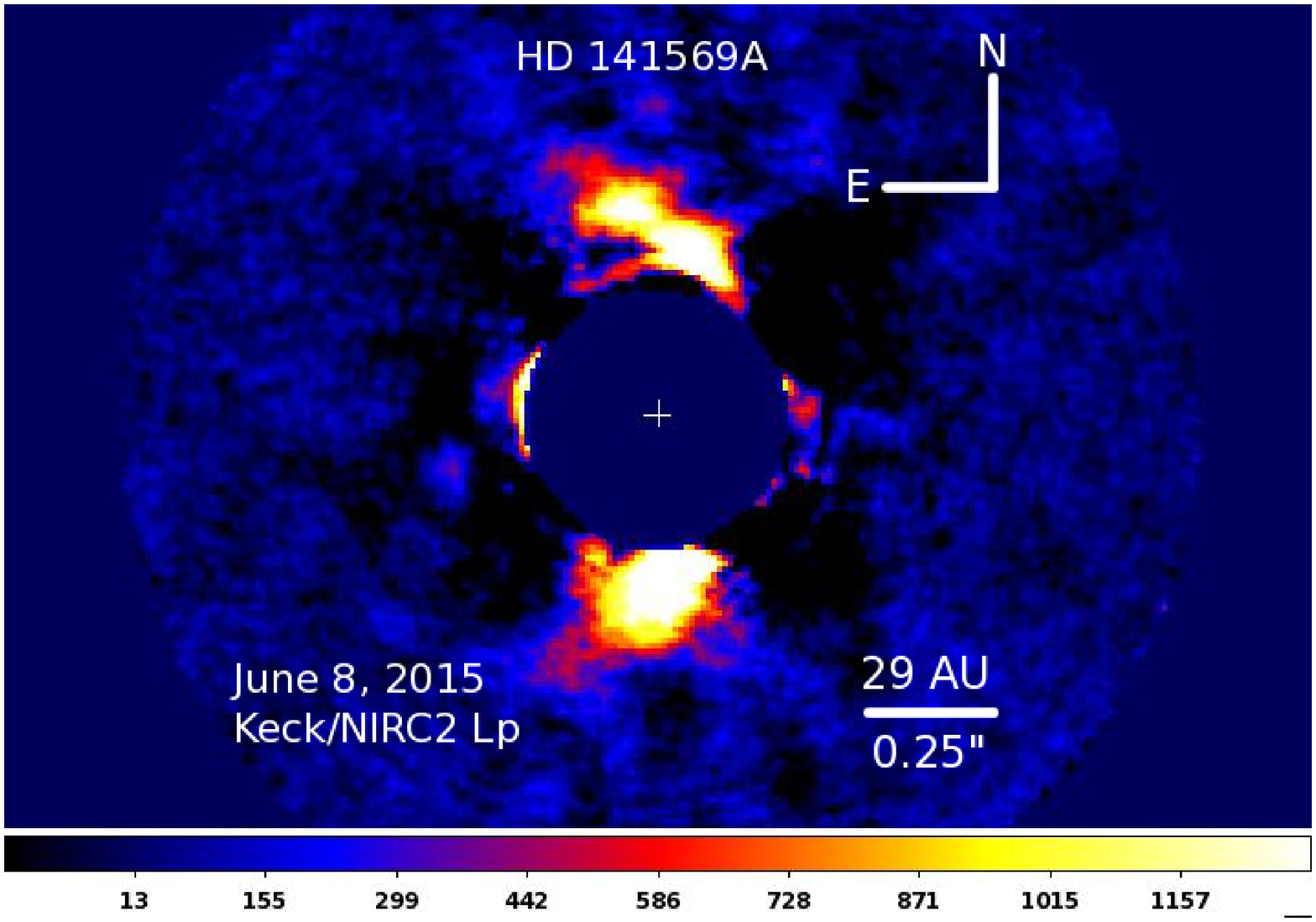}
\includegraphics[scale=0.4,trim=32mm 1mm 32mm 1mm,clip]{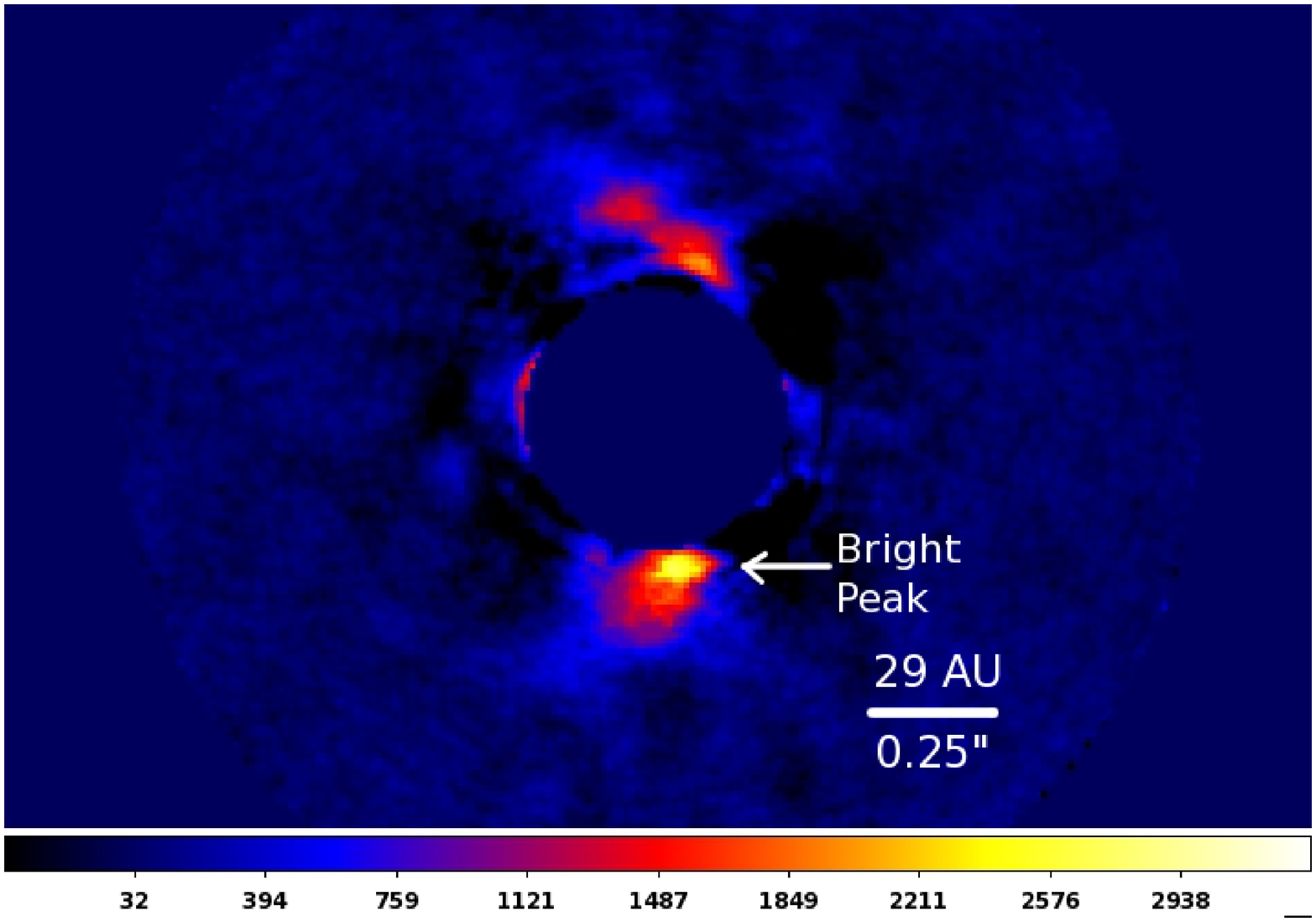}\\
\includegraphics[scale=0.4,trim=32mm 1mm 32mm 1mm,clip]{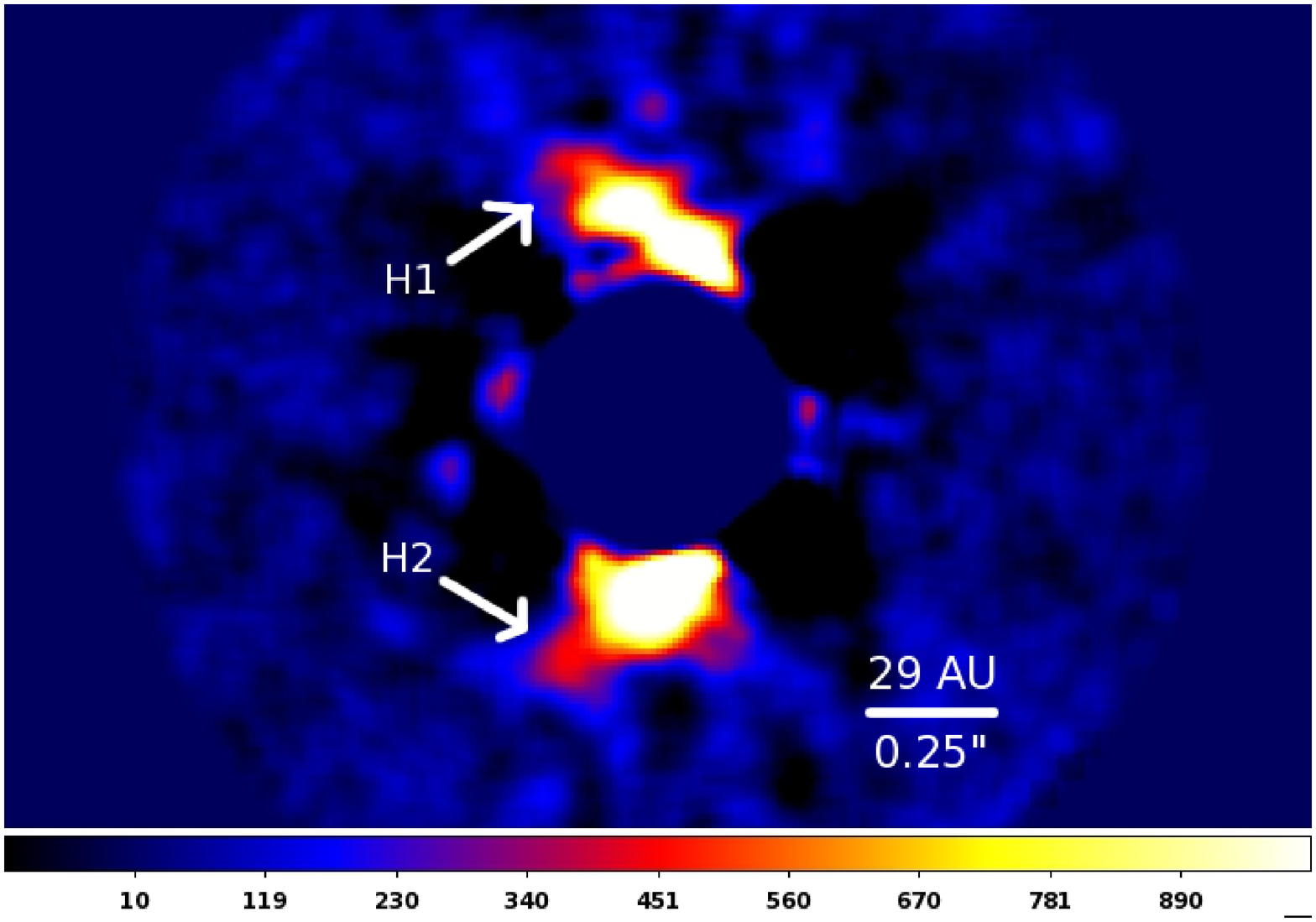}
%\includegraphics[scale=0.4,trim=40mm 1mm 40mm 1mm,clip]{snr.jpeg}\\
%includegraphics[scale=0.4,trim=40mm 1mm 40mm 1mm,clip]{klipkeck.jpeg}
\includegraphics[scale=0.42,trim=32mm 2mm 32mm 10mm,clip]{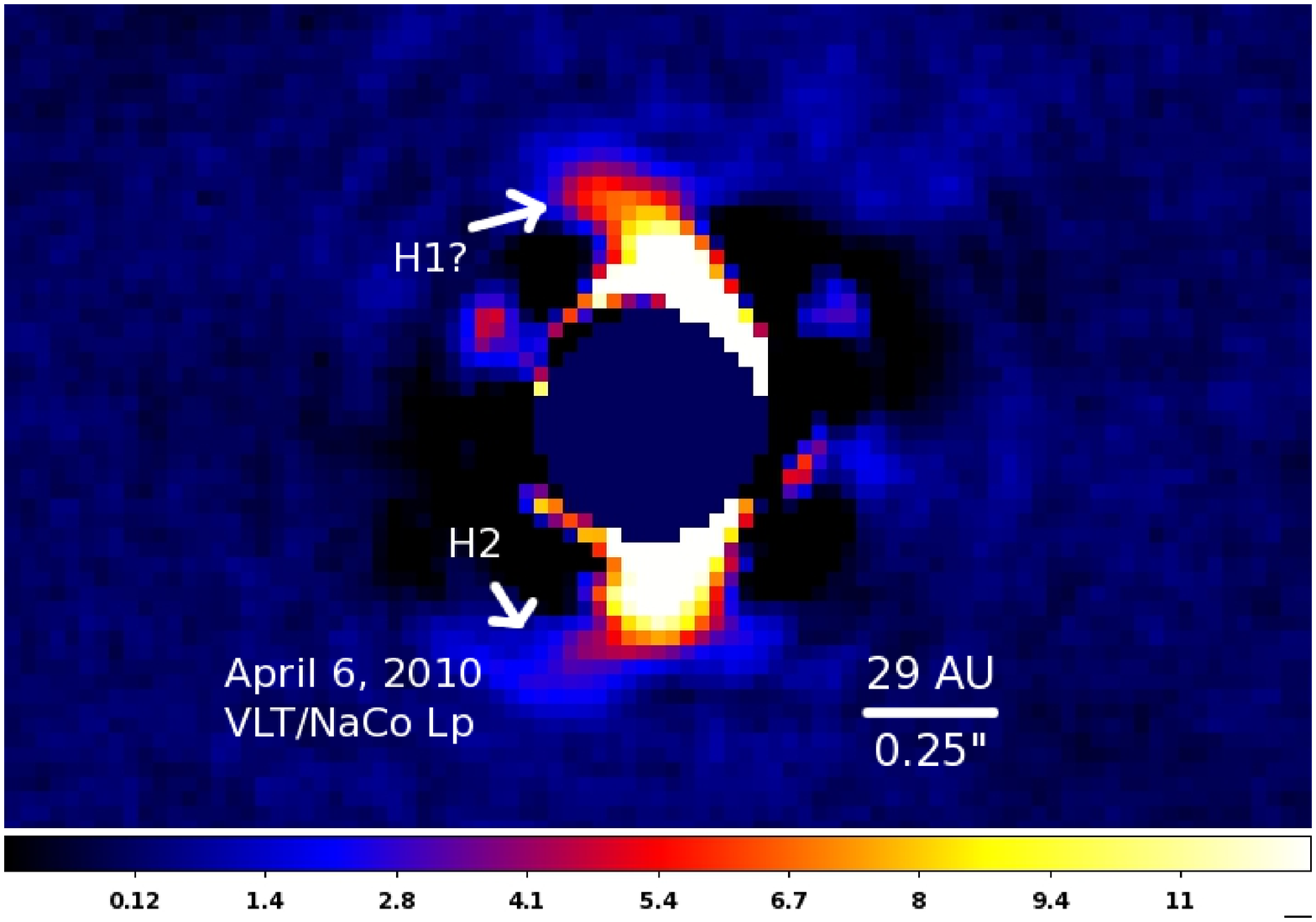}
\caption{Detection of the HD 141569A inner disk:  (top-left) Keck/NIRC2 image with a nominal color stretch (the star's position is identified as a cross), (top-right) a less aggressive color stretch, showing a point-source-like disk bright peak at $r$ $\approx$ 0\farcs{}28, (bottom-left) box-car smoothed image with a hard color stretch, more clearly showing the ``hooks" of emission (H1/2), and (bottom-right) VLT/NaCo image recovering the main disk and the ``hook"-like features.  
  The scale (horizontal bars) is in raw counts.  }
\label{images}
\end{figure}

% Debris Disk}
Figure \ref{images} shows the reduced, combined Keck/NIRC2 image with a nominal image stretch (top-left), a higher dynamic range (top-right), and box-car smoothed to better reveal low intensity extended emission (lower-left) along with the VLT/NaCo image (lower-right).  The Keck image identifies a bright torus-shaped emission, largely on the west side, between 0\farcs{}25 and 0\farcs{}55: a feature recovered by the NaCo data.  Our data do not recover the nested debris disk-like rings nor the extended halo \citep{Konishi2016}\footnote{The high $L^\prime$ sky background likely precludes detecting the low surface brightness outer two rings.      We do not detect the inner disk in existing conventional AO near-IR data because of their low Strehl ratios.}.  However, the newly-identified inner disk appears to have a similar north-south orientation and inclination.   

The data reveal a bright peak at a separation of $r$ $\sim$ 0\farcs{}28 (32 AU) on the south side (top-right panel) that appears point-source-like with a brightness (subtracted from the surrounding disk) comparable to that expected for a 5 $Myr$ old, 5--6 $M_{J}$ planet \citep[][]{Baraffe2003}.  However, dust scattering properties may also explain this peak (see \S 4).     ``Hook-like" features extend from both disk ansae, and are especially visible on the south side in the smoothed image, somewhat similar to the thermal IR-bright arm in HD 100546's disk \citep{Currie2014c}.  

To conservatively define the signal-to-noise ratio per resolution element (SNRE), we replace each pixel with the sum of values enclosed by a FWHM-wide aperture, estimate the radial noise profile of this summed image and divide the summed image by the noise profile.  This procedure yields SNRE $\sim$ 4--6 along the visible trace of the disk in the NIRC2 image between $r$ $\sim$ 0\farcs{}27 and 0\farcs{}55 and slightly lower SNRE at these separations in the NaCo image. 

 From inspection, the disk signal and its self-subtraction footprints, not residual speckles, dominate the pixels at $r$ $\sim$ 0\farcs{}27--0\farcs{}55, thereby biasing the estimate of the noise profile and yielding an underestimated disk SNRE \citep[see also][]{Currie2015b,Thalmann2014}.   Masking a rectangular ``evaluation region" with dimensions 0\farcs{}54 by 1\farcs{}08 centered on the star and defining the radial noise profile from pixels outside this region, we derive a disk SNRE in the Keck image to $\sim$ 8--10 at most separations.    The ``hook-like" (spiral?) features are likewise statistically significant (SNRE $\sim$ 3--5).    We nominally adopt the former (hereafter ``conservative" SNRE) estimate in our disk geometry analysis (\S 4.1) and use the latter (hereafter ``optimistic" SNRE) as our starting point for our disk scattered light forward-modeling (\S 4.2), although these choices does not consequentially affect our results.

 %Thus, we detect the disk at a high SNR in the collapsed image and a high-enough SNR in each spectral channel to obtain the disk surface brightness as a function of wavelength.

\section{Analysis}
%Deriving debris disk parameters from data sets using advanced PSF subtraction techniques is a challenging problem \citep[e.g.][]{Esposito2014,Mazoyer2014}.  
To derive the HD 141569A inner disk geometry, we follow the same approach used for analyzing HD 115600's disk \citep{Currie2015a}.  First, we derive the disk's basic geometry from ellipse fitting.  Second, we use forward-modeling to fine-tune these properties and calculate second-order properties of the disk (e.g. scattering function), assuming that we are seeing optically-thin, scattered light emission.  We focus our analysis on the higher-quality Keck/NIRC2 data.
% using our model of the disk geometry. 
\subsection{Geometry}
From 
 the IDL \textit{mpfitellipse} package we first define a trace of the disk, where the pixels are weighted by their conservative SNRE.   Second, we constructed a grid of ellipse parameters around the best-fit set determined by \textrm{mpfitellipse}, calculating a value using the ``maximum merit" procedure \citep{Thalmann2011}.  We repeat this step using different ranges in radii and different cutoffs in SNRE for the disk trace (e.g. SNR $>$ 3, 5; $r$ = 0\farcs{}25--0\farcs{}5, 0\farcs{}25--0\farcs{}55) to define best-estimated values and associated uncertainties.  

The disk geometry generally agrees well with the most precise estimates for the outer disks' geometry \citep{Mazoyer2016,Konishi2016}.  We derive a best-fit position angle of PA = -1.2$^\circ$ $\pm$ 2.4$^\circ$.  While we derive an inclination of $i$ = 56$^\circ$ $\pm$ 4$^\circ$ considering the mean value of all estimates, a large subset of solutions center around 60$^{o}$.    The disk semimajor/minor axes are 0\farcs{}340 $\pm$ 0\farcs{}020 ($\sim$ 39.1 AU $\pm$ 2.2 AU) and 0\farcs{}189 $\pm$ 0\farcs{}010 (21.9 $\pm$ 0.2 AU), respectively.  
%The semimajor axis and visible extent of the disk ansae are within the range of radii encompassing the current Kuiper belt; the luminosity-scaled ($r_{major}$/$\sqrt{L_{\star}}$) semi major axis (extent of the ansae) are $\sim$ 22 AU (17--25 AU), comparable to the predicated distances of major models of the early, pre-stirred Kuiper belt,: e.g. the Nice model  \citep[][]{Levison2008} or \citet{Nesvorny2015}.   
The projected disk center is offset from the star is $\Delta$x,$\Delta$y = -0\farcs{}044 $\pm$ 0\farcs{}016 (6.1 $\pm$ 1.3 AU), 0\farcs{}014 $\pm$ 0\farcs{}010 (1.6 $\pm$ 1.1 AU).
%, broadly consistent with the northwest offset of the inner ring \citep{Biller2015}.
%and the peak signal of the two ansae differ by $\sim$ 0.5 pixel in angular separation ($\sim$ 1 AU).

\subsection{Disk Forward Modeling}
To infer additional disk properties, we generate a grid of synthetic scattered light images using GRaTeR \citep{Augereau1999} and forward-model these synthetic disks through A-LOCI to compare the processed model disk image with the real disk image \citep{Esposito2014}.
 %We insert a disk model into a sequence of empty images with position angles identical to those of our science sequence, and convolve the model in each spectral channel with the appropriately-sized PSF.     We performed PSF subtraction on the images containing the model disk using the same A-LOCI coefficients that were applied to the real data.  

Table \ref{diskmodels} summarizes the model parameter space.  For simplicity, we adopt the position angle determined from our ellipse modeling (-1.2$^\circ$).  We consider a nominal inclination of 56$^{o}$ and an inclination of 60$^{o}$ favored by a subset of our ellipse-fitting results.  
%The ansae of the disk along the major axis have a small measured offset from the center of the star.   Additionally, 
%We defined the argument of pericenter to be along% the visible disk major axis:  departures from these values resulted in brightness asymmetries between the disk ansae inconsistent with the data.   
We tuned our parameter search to focus on addressing specific questions about the HD 141569A morphology.     First, to assess whether the disk (made visible by small, scattering dust grains) coincides with the gas distribution or is radially segregated, we consider photocenters of 36.9, 39.1, and 41.3 AU (consistent with our ellipse modeling) and a photocenter at 25 AU: roughly the separation corresponding to the half-maximum of the CO gas \citep{Goto2006}.   Second, we assess whether the emission originates from a sharp debris ring with a steep drop in density away from the disk photocenter ($\alpha_{in}$, $\alpha_{out}$ =10, -10), the visible edge of a continuous distribution that (when annealed by processing) only ``appears" to be truncated ($\alpha_{in}$ = 1), or a dust torus (not a sharp ring) with an intermediate power law decay at separations interior to the photocenter ($\alpha_{in}$ = 2.5--5) (see Augereau et al. 1999 for definitions).  We consider power law decays exterior to the disk photocenter of  $\alpha_{out}$ = $-2.5$ to $-10$, varying, Henyey-Greenstein scattering parameters $g$ (0--0.2), the disk offsets from the star in both x and y, and disk scale heights ($ksi_{o}$ = 3--5 AU).
%7.5, 10; $\alpha_{out}$ = $-$5, $-$7.5),  the disk eccentricity (0--0.3), and the offset along the major axis (0 and 1 AU).     
%While our parameter space search is not exhaustive, values outside these ranges (e.g. $g$ $>$ 0.15) yielded processed synthetic disk images strongly discrepant with the real data (e.g. models with $g$ $>$ 0.2 yield inaccurate brightness ratios between the disk ansae and other regions).
%regions at smaller angular separations.
  
%To match the observed brightness we fix the
%parameters in the model and scale the flux and define the fit of the model to the data using the real and synthetic collapsed images convolved by the PSF as in our SNR calculations.   
To identify the best-fitting disk models, we closely follow the methods from \citet{Thalmann2014}.  Briefly, we bin down the Keck image, the model image, and the noise profile to the Keck/NIRC2 spatial resolution to compare the data and model at effectively independent data points and compute $\chi^{2}$ from the residuals of the binned image over the angular separation where the disk detection is significant and negligibly contaminated by residual speckles ($r$ $\sim$ 0\farcs{}27--0\farcs{}55).    As noted in \S 3, determining a radial noise profile (and thus a robust SNRE for the disk) is extremely difficult due to biasing from the disk and self-subtraction footprints, impeding our ability to quantify a robust estimate of the absolute goodness of fit for the models.   Thus, we iteratively rescale the radial noise profile from the ``optimistic" SNR map such that the best-fitting model has $\chi^{2}_{\nu}$ = 1 and focus simply on the family of best-fitting models: i.e. those fulfilling $\chi^{2}$ $\le$ $\chi^{2}_{min}$ + $\sqrt{2\times N_{data_{binned}}}$ \citep{Thalmann2013}, which for our case implies $\chi^{2}_{\nu}$ $\le$ 1.124.  

\begin{figure}
\centering
%\epsscale{0.8}
%\includegraphics[scale=0.3,trim=40mm 1mm 40mm 1mm,clip]{alocikeck.jpeg}
%\includegraphics[scale=0.3,trim=40mm 1mm 40mm 1mm,clip]{alocikeck.jpeg}
%\includegraphics[scale=0.3,trim=40mm 1mm 40mm 1mm,clip]{alocikeck.jpeg}\\
\includegraphics[scale=0.3,trim=32mm 1mm 32mm 1mm,clip]{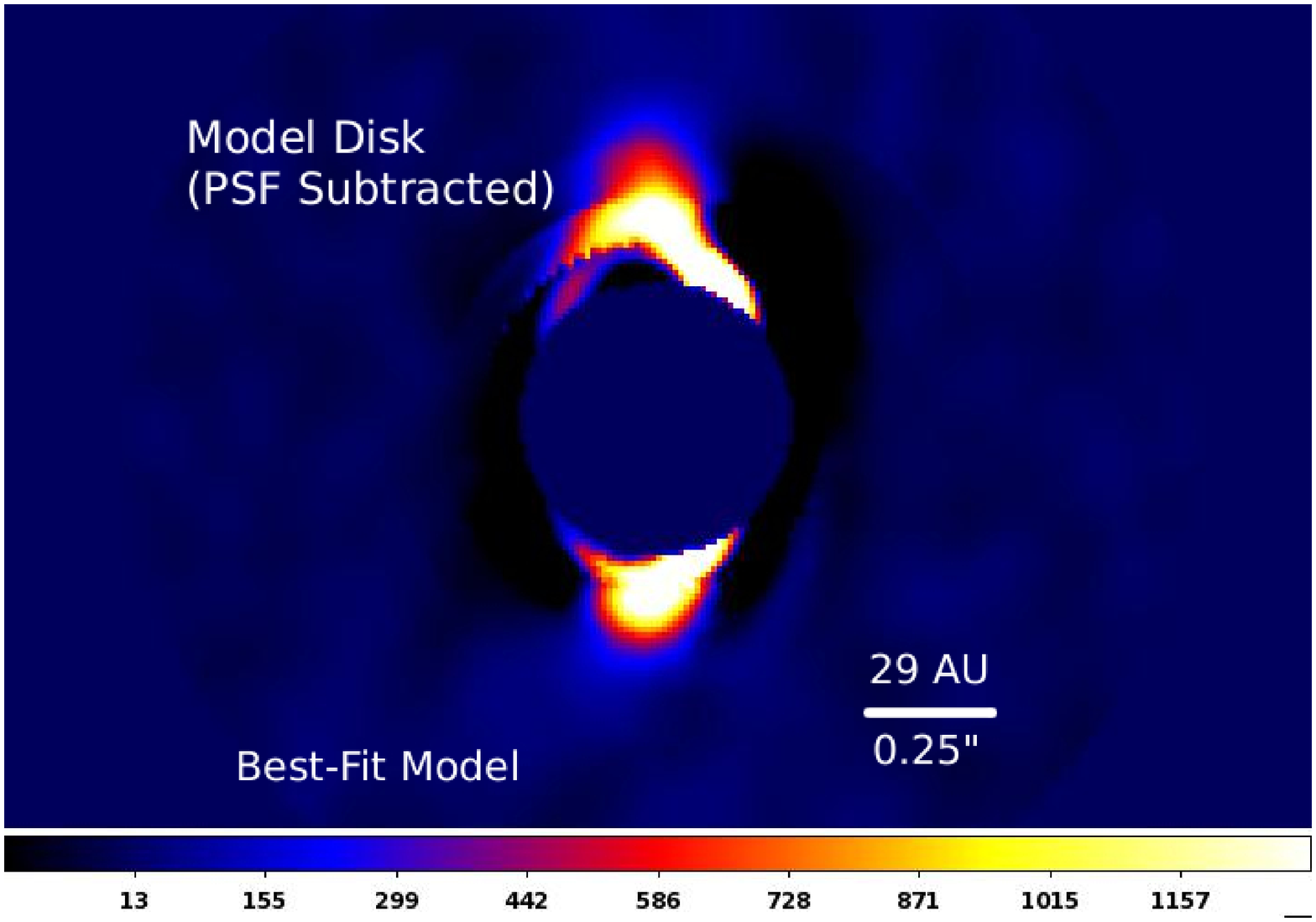}
\includegraphics[scale=0.3,trim=32mm 1mm 32mm 1mm,clip]{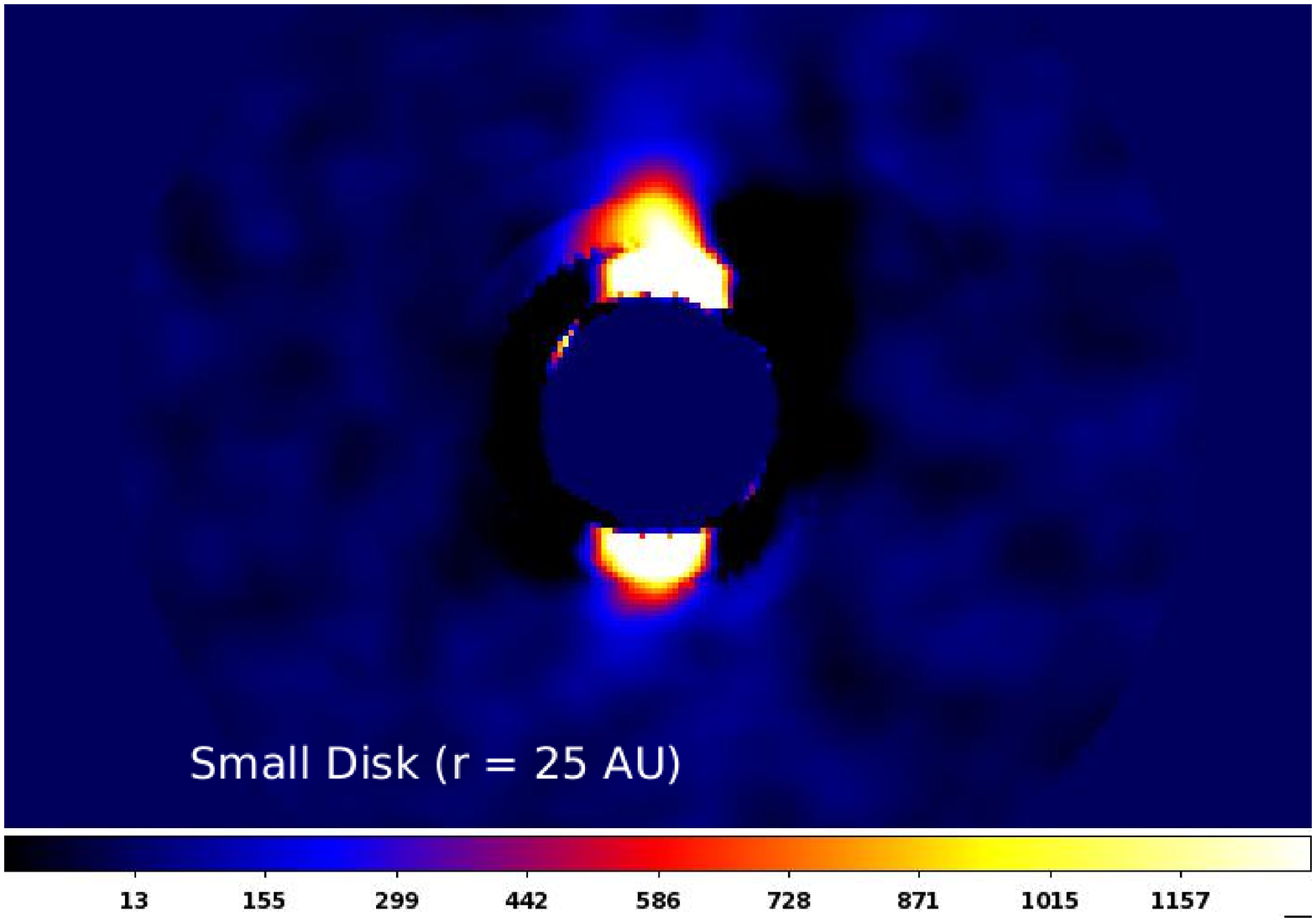}
\includegraphics[scale=0.3,trim=32mm 1mm 32mm 1mm,clip]{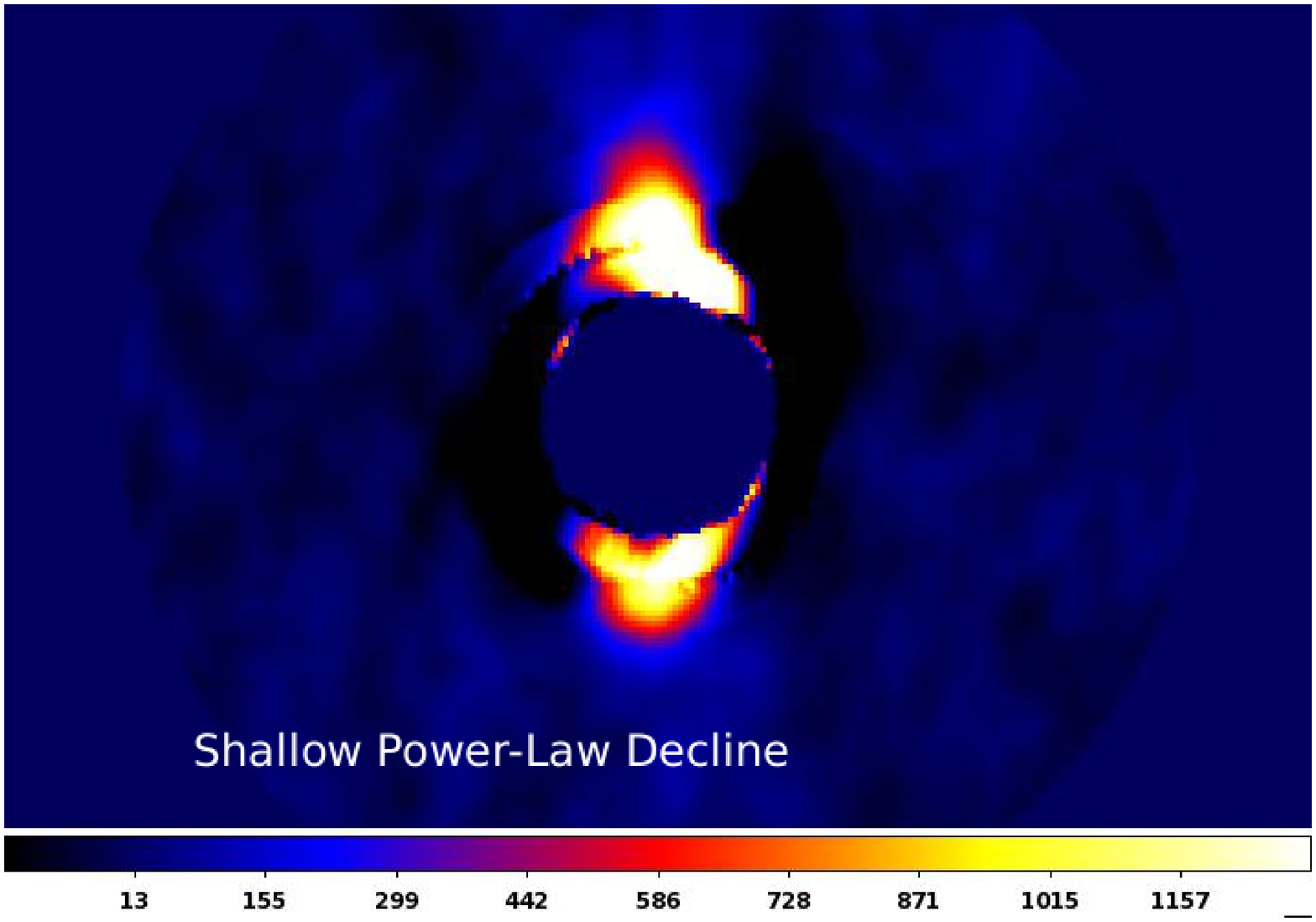}\\
\includegraphics[scale=0.3,trim=32mm 1mm 32mm 1mm,clip]{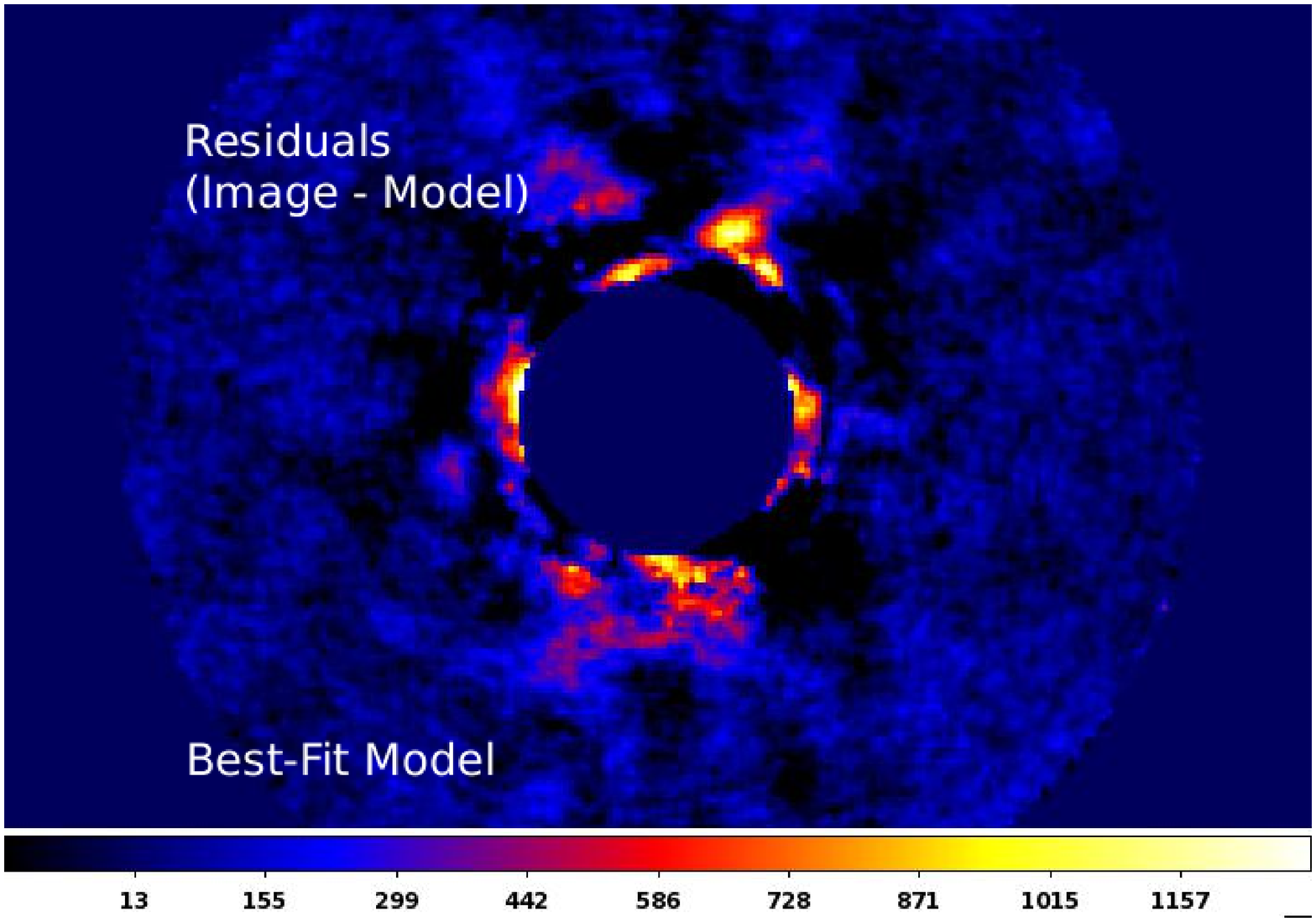}
\includegraphics[scale=0.3,trim=32mm 1mm 32mm 1mm,clip]{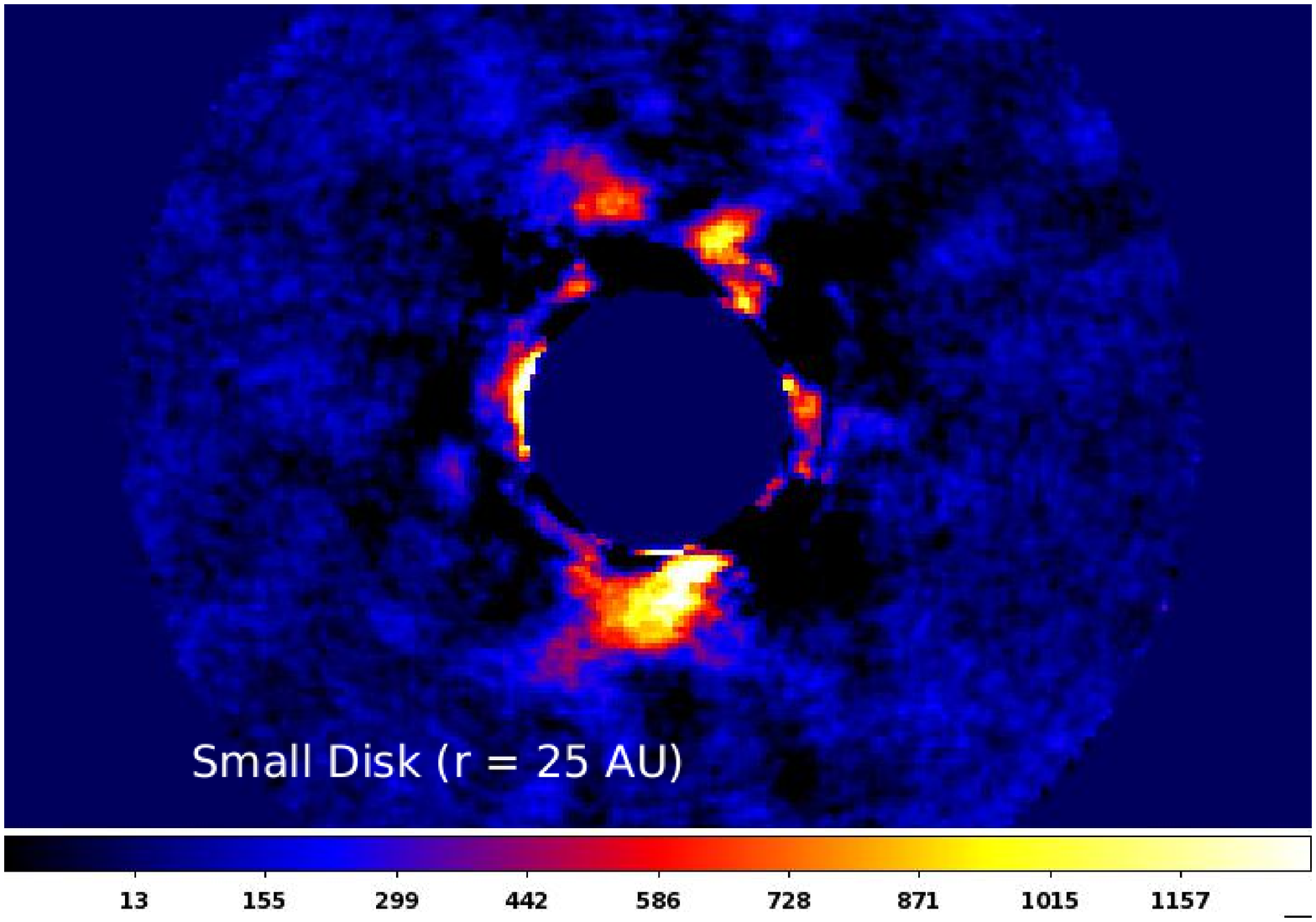}
\includegraphics[scale=0.3,trim=32mm 1mm 32mm 1mm,clip]{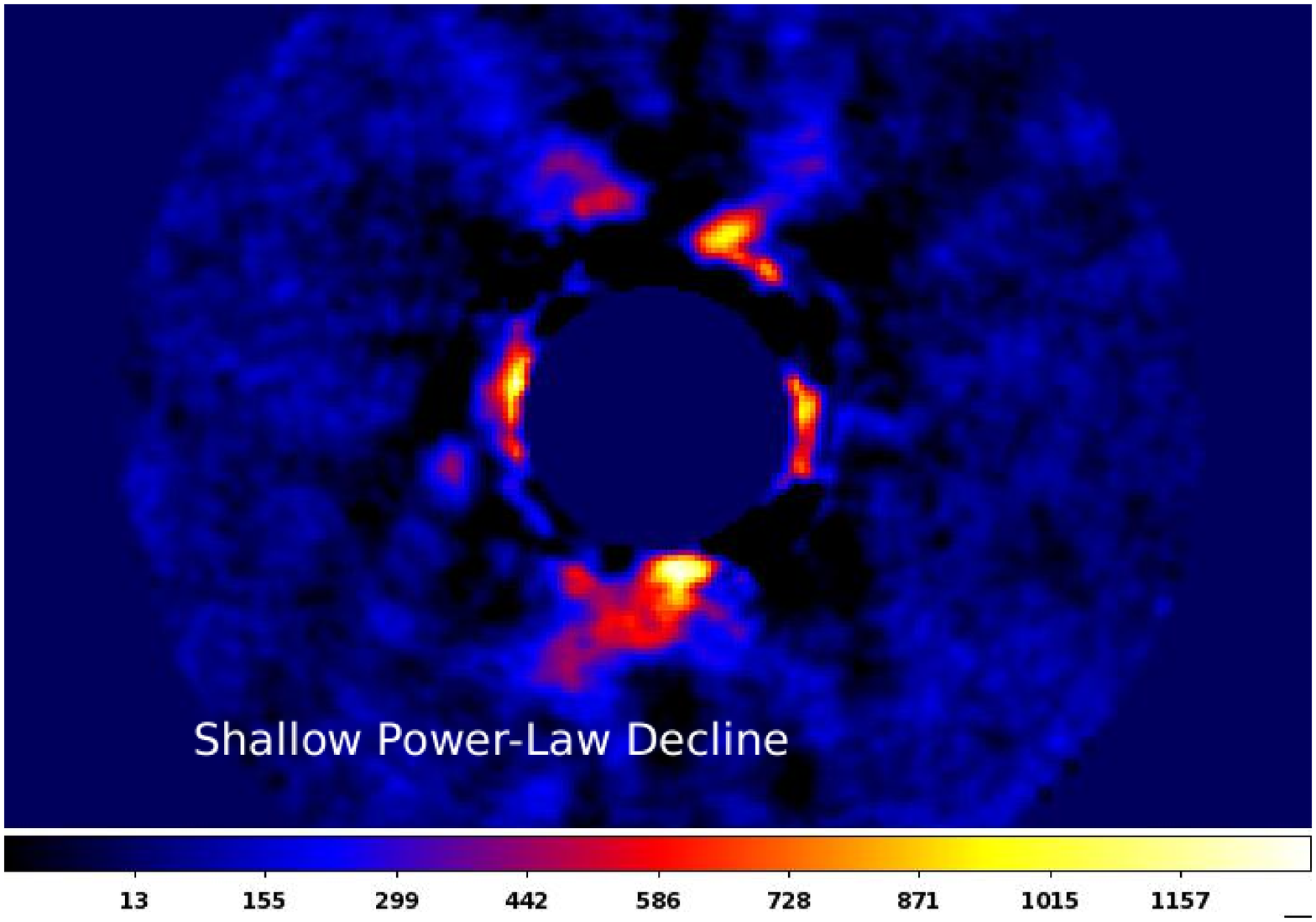}
\caption{Forward-modeling of the inner disk.  (top rows) PSF-subtracted models: (left) best-fit model -- $g$ = 0, $\alpha_{in}$ = -5, $\alpha_{out}$ = -10, $r_{o}$ = 39.1 AU, PA = -1.2$^{o}$, $i$ = 56$^{o}$, $ksi_{o}$ = 5 AU ($\chi^{2}_{\nu}$ = 1); (middle) best-fitting model with $r_{o}$ = 25 AU ($\chi^{2}_{\nu}$ = 1.31); and (right) best-fitting model with a shallow ($\alpha_{in}$ =1) power-law decline in density interior to the ring photocenter ($\chi^{2}_{\nu}$ = 1.11).  (bottom rows) Residuals of each model subtracted from the Keck/NIRC2 image.  The bright point source-like peak is removed in our best-fit model, but the hook-like features remain.  Models with small $r_{o}$ more similar to the CO gas peak can be ruled out, and those lacking a highly-evacuated region interior to the disk photocenter are disfavored.
%   -- produced by GRATER, the best-fit model subtracted after processing (middle), and residuals of the model's subtraction from our target image (right).  (middle rows) Same as the top row except for a model with a very large $\alpha_{in}$ value consistent with a sharp ring-like disk, and (bottom rows) same but for a disk whose photocenter is at $r$ = 25 AU, the location of the CO gas peak found in \citet{Goto2006}.  
%The HD 141569A inner disk is not best reproduced by a ring-like structure and identifies dust whose peak emission lies beyond the CO gas peak.  
}
\label{forwardmodel}
\end{figure}

The best-fit model accurately reproduces the disk morphology (Figure \ref{forwardmodel}) and, when subtracted from the Keck image, nulls its signal, including the point source-like peak, but leaves the hook-like features largely intact (SNRE $\sim$ 3.5--4.5 in the residual image).   The integrated signal of the best-fit model is $\approx$ 20 mJy.  Its surface brightness along the major axis (i.e. the model shown in the top-left panel prior to signal loss from PSF subtraction) ranges from $\approx$ 7.5 mag arcsec$^{-2}$ at the photocenter (39 AU) to $\approx$ 10 mag arcsec$^{-2}$ at the widest separations where it is detected robustly ($r$ $\sim$ 0\farcs{}55).  

 We can decisively rule out some model phase space and identify key trends.   First, models with a photocenter of $r$ = 25 AU (or, by extension, at smaller separations) are inconsistent with the data, yielding especially high residuals on the south side (Figure \ref{forwardmodel}, middle panels). 
% Thus, we can easily rule out a scattered light dust distribution tracking that of the CO gas \citep[25 AU,][]{Goto2006}.  
Models with $\alpha_{out}$ = -2.5 are ruled out: the disk requires a sharp density cutoff exterior to the photocenter.      The dust is also likely (near-)neutral scattering.

While our analysis does not formally preclude models with weak depletion interior to the photocenter ($\alpha_{in}$ = 1), even the best-fit of these models (right panels) is marginally acceptable, yielding clearly higher residuals.    The $\chi^{2}$ distribution for $\alpha_{in}$ = 2.5 -- 5 is systematically skewed towards smaller values, indicating a better fit: models with these power laws dominate the family of best-fitting solutions.  
\textit{Thus, our disk modeling favors a torus of neutral scattering dust at $\approx$ 36--41 $AU$
with sharp truncation at larger disk radii and significant depletion at smaller radii. This material largely lies outside the CO gas concentrated at 11--25 $AU$}.
%\textit{Thus, our disk forward-modeling favors neutral-scattering dust at a larger separation than found for CO gas and is confined to a torus: sharply truncated exterior to 39 $\pm$ 2.2 AU and heavily depleted interior to this radius}. 

\subsection{Limits on Planets}
To place limits on the presence of unseen planets that may be perturbing the inner disk, we reprocess the data using aggressive A-LOCI settings (e.g. $\delta$ = 0.5).
%\footnote{Formally, to identify the set of parameters yielding the deepest contrast, we would have to inject synthetic point sources into the images over an n-dimensional parameter grid or Markov Chain Monte Carlo simulation.  But the likely contrast gain with optimization ($\sim$ 2-3x at best) is small compared to the huge dynamic range in predicted contrast at/near 1 $M_{J}$. }  
We determine the throughput loss-corrected radial noise profile 
 \citep{Lafreniere2007a,Marois2008}.  The \citet{Baraffe2003} hot-start evolutionary models allowed us to map between $L^\prime$ brightness and planet mass.  
 
While previous analysis suggests that HD 141569 A is $\sim$ 5 $\pm$ 3 Myr old, we reinvestigated its age, comparing the HR diagram positions of the primary to \citet{Dotter2008} models.  
%Using the \citet{Baraffe1998} models, HD 141569 BC are both $\approx$ 5 Myr old, but the \citet{Dotter2008} evolutionary tracks imply a coeval age of less than $\approx$ 3 Myr.  In %contrast, 
HD 141569 A appears too low in luminosity to be consistent with a 5 Myr age.  It's placement on an HR diagram implies an age between 6 Myr and the zero-age main sequence (starting at $\approx$ 10 Myr).  For our planet mass limits, we consider the nominal age and an older, revised age of 7.5 Myr\footnote{Conversely, the placement of the M star companions on the HR diagram imply a \textit{younger} age.   The discrepancies of pre-main sequence tracks at high vs. low masses is beyond the scope of this paper.}.
% revised ages.
%$a nominal age of 5 $Myr$ and the older, best-estimated revised age.

%.  Thus, we consider results for 
%We split our parameter search in two: focusing on regions at/interior to the inner disk in projected separation and then exterior to our outer working angle ($r$ $\sim$ 1\farcs{}1).    Previous experience with HR 8799 $L^\prime$ observations of similar depth suggests that the contrast in the latter region will be less sensitive to algorithm set up, leading us to consider only a sparse grid (e.g. $\Delta$$\delta$, $N_{A}$ = 0.1, 10).  

\begin{figure}
\centering
%\epsscale{0.8}
%\includegraphics[scale=0.4,trim=40mm 1mm 40mm 1mm,clip]{alocikeckhighdyn2.jpeg}
%\includegraphics[scale=0.4,trim=40mm 1mm 40mm 1mm,clip]{alocikecklowdyn2.jpeg}
\includegraphics[scale=0.6]{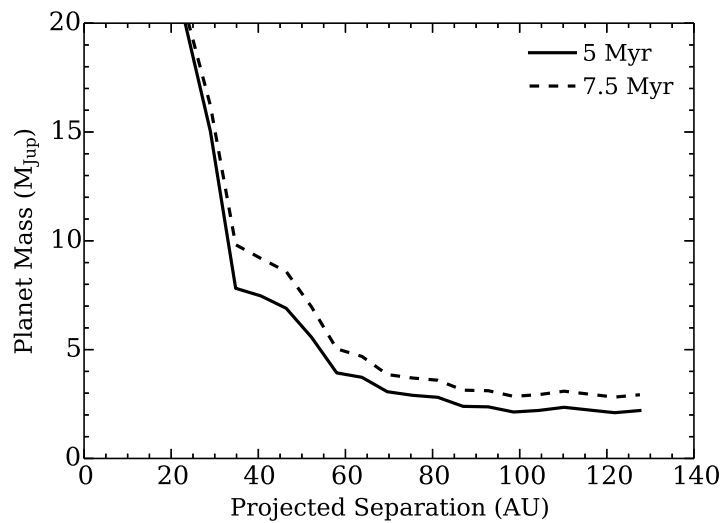}
\caption{
%(left) A revised age estimate for HD 141569A based off of the HR diagram positions of its two M dwarf companions using the \citet{Dotter2008} isochrones. (right) 
Planet detection sensitivities assuming the \citet{Baraffe2003} hot-start evolutionary models and a nominal age of 5 Myr (solid line) or a proposed revised age of 7.5 Myr (dashed lines) from an aggressive A-LOCI re-reduction of the Keck/NIRC2 data.  %Our data can rule out planets with masses just slightly above that of Jupiter exterior to the inner disk.
%A-LOCI processed image with a less aggressive color stretch, showing a strong, point-source like disk bright spot at $r$ $\approx$ 0\farcs{}27.  (right) Same image but box-car smoothed with a hard color stretch, more clearly showing the ``hook" of emission wrapping clockwise from due south.
%Detection of the HD 141569A inner disk:  (top-left) reduced Keck/NIRC2 image processed with A-LOCI, (top-right) signal-to-noise ratio map of this image, (bottom-left) KLIP-processed Keck/NIRC2 image, and (bottom-right) A-LOCI processed VLT/NaCo archival image.  The HD 141569A inner disk appears as a bright torus of emission extending to $r$ 0\farcs{}045 in the Keck images with a SNR of $\sim$ 5--20.  The NaCo image recovers this torus and the ``hook"-like feature due south that may be a spiral arm.  For the images, the scale (horizontal bars) is in counts.  The SNR map color stretch is in units of $\sigma$}
}

\label{planetlimits}
\end{figure}

At $r$ = 0\farcs{}25--0\farcs{}5 (29--45 AU), the azimuthally-averaged sensitivity limit is roughly 7.5--10 $M_{J}$ (Figure \ref{planetlimits}), shallow since the disk's bright signal substantially drives up the residual noise estimate.  Exterior to the inner disk, our sensitivity curve excludes planets with masses of 2--3.5 $M_{J}$ (3--4 $M_{J}$) at projected separations of 60--120 AU assuming an age of 5 (7.5) $Myr$.   If the planet luminosity evolution is better described by a cold start model, our mass limits are significantly poorer.
%Thus, an externally-perturbing planet likely has to be equal or lower in mass than the least massive directly-imaged planets \citep[e.g. GJ 504 b;][]{Kuzuhara2013}, while a planet located within the disk could be comparable in mass to ROXs 42Bb or $\beta$ Pic b \citep[9--10 $M_{J}$;][]{Currie2014a,Lagrange2010}.

%Debris disk dynamical modeling sets stronger limits on unseen planets than from the GPI contrasts limits alone \citep[contrast $>$ 10$^{-5}$ at r $<$ 0\farcs{}4 (44 AU) or $M_{\mathrm{planet}}$ $>$ 7 $M_\mathrm{J}$; c.f.][]{Baraffe2003}.
%\subsection{Spectral Analysis}

\section{Discussion}
\begin{figure}
\centering
\includegraphics[scale=0.25]{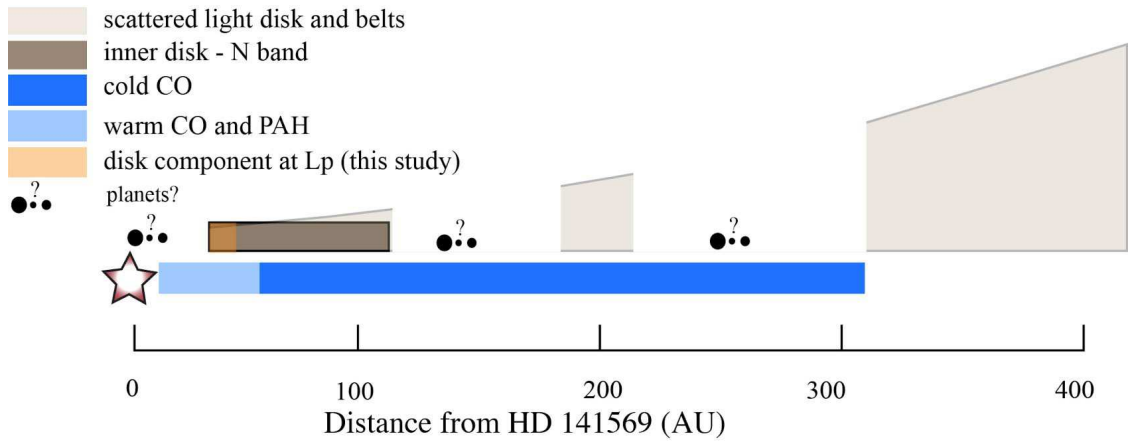}
\caption{The different circumstellar dust and gas components of HD 141569A \citep[after ][]{Pericaud2014}: the two outer rings of dust first resolved by HST, the inner torus of dust (this work), extended 10--20 $\mu m$ emission \citep{Fisher2000,Marsh2002}, the inner dust halo \citep{Konishi2016}, and multiple reservoirs of CO  gas.}
\label{cartoon}
\end{figure}

Figure \ref{cartoon} depicts an updated schematic of the complex HD 141569A circumstellar environment.  The system now includes three scattered light-resolved, concentric rings/torii of material spanning between 25 and 500 $AU$, a halo of small dust emission located between the innermost and middle ring, and significant localized structure in the middle and outermost rings plus multiple (candidate) spirals/hooks in all three rings indicative of perturbations by stellar/substellar companions \citep[this work,][]{Konishi2016,Mazoyer2016}.    Interspersed between these dust components are multiple gas reservoirs \citep{Goto2006,Thi2014}.   

Furthermore, the inner disk dust and gas are likely not well mixed: dust is concentrated in a torus with a larger radius than the CO gas and is likely heavily depleted at smaller separations, implying that its dust cavity is larger than the gas cavity.   
Recent analysis of ALMA data by \citet{vanderMarel2015,vanderMarel2016} likewise provides strong evidence for radial segregation of warm dust and gas in transitional disks, with dust cavities up to 3 times larger and an order of magnitude more depleted than gas cavities.   However, the disks they analyzed are generally massive, optically thick in the mid-infrared, and vigorously accreting.  HD 141569 A has weak/negligible accretion, is optically thin in the mid-infrared \citep{Marsh2002}, and low mass.  

While the heavily depleted dust cavities in canonical transitional disks are consistent with the planet trapping scenario \citep{Pinilla2012,vanderMarel2013,vanderMarel2016}, HD 141569 A's radially-segregated disk may be shaped by other mechanisms.    For instance, the dimensionless Stokes parameter in the Epstein drag regime, appropriate for small grains, St 
$\sim$ $\frac{\pi \rho_{s}a}{2\Sigma_{gas}}$
%$\sim$ \frac{\pi\rho_{s}a}{2\Sigma_{gas}} 
\citep[c.f.][]{Weidenschilling1977,Birnstiel2010}, is $\ll$ 1 for 1--100 $\mu m$-sized grains in typical, optically thick protoplanetary disk conditions: such dust is well mixed 
with the gas.  However, as the gas density drops, smaller grains (more easily probed by scattered light observations) decouple from the gas.   
%Adopting the gas number density of 10$^{21}$ cm$^{-2}$ ($\Sigma_{gas}$ $\sim$ 0.004 g cm$^{-2}$) at the CO gas ring's peak from \citet{Thi2014} and assuming porous (1.25 g cm$^{-3}$) grains, dust that is 4--20 $\mu m$ should be indeed (marginally) decoupled from the gas (St $\approx$ 0.2--1). 
If the 4-20 $\mu m$ dust particles, accessible by our observations, are indeed (marginally) decoupled from the gas (St $\approx$ 0.1-1), the gas surface density should be of the order $\approx$ 0.004 g cm$^{-2}$, assuming porous grains (1.25 g cm$^{-3}$). This is consistent with the gas number density found by \citet{Thi2014} (10$^{-21}$ cm$^{-2}$).
 Under the influence of radiation pressure, dust grains of these sizes in optically-thin disks can be pushed outwards and form ring/torus-like structures \citep{Takeuchi2001}.
%Thi et al. 2014 (10^21 cm^-2).

%If the warm dust emission we identify is a torus of emission radially segregated from the gas and created by some trapping/radial drift mechanism, 
The morphology of the newly-discovered inner disk as probed by larger grains could be more striking.   Since longer wavelength observations better probe large grains, the disk's presence could be recovered and its structure clarified by high spatial resolution mm interferometry.  Its measured submillimeter dust continuum emission \citep[$\approx$ 8.2 $mJy$;][]{Flaherty2016} is easily within detectability with ALMA;  mid-IR interferometry with VLTI/Matisse may also resolve its emission at high resolution.

%Until recently, 
%While 10--20 $\mu m$ imaging provided evidence for and marginally resolved a warm, solar system-scale dust component previously undetected in scattered light, our paper for the first time images emission on this scale with high spatial resolution 

%and should a prime new target for follow-up planet searches with GPI or SPHERE: though not with GPI because they are children.  
% with rather neutral-scattering dust

%\textbf{Acknowledgements} --
\acknowledgements 
\textbf{Acknowledgements} --  
We thank Scott Kenyon and Mengshu Xu for helpful comments.
%and Prof. Larry Kimura of the Hawaiian Language Studies program at the University of Hawaii-Hilo for verifying the proper spelling/pronunciation of Maunakea.
CAG is supported under the NASA Origins of Solar Systems program NNG13PB64P.
We wish to emphasize the pivotal cultural role and reverence that the summit of Maunakea has always had within the indigenous Hawaiian community.  We are most fortunate to have the privilege to conduct scientific observations from this mountain.    
%Finally, we thank Prof. Larry Kimura of the College of Hawaiian Language at the University of Hawaii-Hilo for confirming that Maunakea (or potentially Mauna A Wakea), not "Mauna Kea", is the more historically and linguistically accurate rendering.
%We thank Scott Kenyon, and Mengshu Xu for helpful conversations.   
%We thank the anonymous referee, Wladimir Lyra, Eric Mamajek, and Mengshu Xu for helpful comments; Jean-Charles Augereau for use of GRATER; Fredrik Rantakryo for executing these queue-mode observations; and the GPI Early Science Time Allocation Committee and Gemini Director Markus Kissler-Patig for supporting this program.  

 %Finally, T.C. thanks Mengshu Xu also for unwavering moral support  
 % and for not publicly announcing the name of this star or this result prior to journal submission.  You will get your own ring, soon: not as big as this one but with a higher albedo.

{}

\begin{deluxetable}{lcllccccccc}
\setlength{\tabcolsep}{0pt}
\tablecolumns{4}
\tablecaption{HD 141569A Disk Forward Modeling}
\tiny
\tablehead{{Parameter}&{ Model Range}&{ Best-Fit Model} & { Well-Fitting Models} }
\startdata
\textit{fixed values}\\
$PA$ & -1.2$^{o}$ & " & "\\
\textit{varied values}\\
$i$ & 56--60$^{o}$ & 56$^{o}$ & 56--60$^{o}$ \\
$r_{o}$ (AU)& 25, 36.9, 39.1, 41.3  & 39.1 & 36.9--41.3\\
$\Delta x$  (AU) & 4.3, 6.1, 7.9 & 7.9 & 4.3--7.9\\
$\Delta y$  (AU) & 0.5, 1.6, 2.7 & 0.5& 0.5--1.6\\
$\alpha_{in}$ & 1, 2.5, 5, 10 & 5 & 1--10  \\
$\alpha_{out}$ &$-2.5$, $-5$, $-10$ & $-10$ & $-5$--$-10$\\
$g$ & 0, 0.1, 0.2 & 0 & 0--0.1\\
$ksi_{o}$ (AU) &  3, 5 & 5 & 3--5\\
\enddata
\tablecomments{Range of best fit and well-fitting model parameters as determined by our $\chi^{2}$ criterion (see \S 4).  
%We define the ``best fit" model as the one minimizing $\chi^{2}$ .
%over $i$ convolved pixels where 
%$\chi^{2}$ = $\sum\limits_{i}$ (model$_{i}$-image$_{i}$)$^{2}$/$\sigma_{i}^{2}$.  
%``Good-fitting" models are those satisfying $\chi^{2}$ $\le$ $\chi^{2}_{min}$ + $\sqrt{2\times N_{pixels (binned)}}$.
}
\label{diskmodels}
\end{deluxetable}

\end{document}